\documentclass[12pt]{article}

\usepackage[a4paper, total={6.8in, 10in}]{geometry}

\usepackage{xcolor}
\usepackage{xargs}
\usepackage[hyperrefurl]{research18}
 \usepackage[caption=false,font=footnotesize]{subfig}

\newcommandx{\blue}[1]{\textcolor{blue}{#1}}

\newcommand{\CiRh}{C^{(\textnormal{I})}(\Rh)}

\newcommand{\defn}{\triangleq}

\let\P\relax
\newcommand{\P}{\mathcal{P}}
\newcommand{\M}{\mathcal{M}}
\newcommand{\T}{\mathcal{T}}

\newcommand{\Rh}{R_\textnormal{h}}

\newcommand{\Unif}{\textnormal{Unif}}

\let\I\relax
\newcommand{\I}{\textnormal{I}}

\newcommandx{\strong}[3][2=\epsilon,3=n]{\mathcal{A}^{* (#3)}_{#2} (#1)}
\newcommandx{\weak}[3][2=\epsilon,3=n]{\mathcal{A}^{ (#3)}_{#2} (#1)}
\newcommandx{\typen}[2][2=n]{\top ^{ (#2)}_{#1}}
\newcommandx{\contypen}[3][3=n]{\top ^{ (#3)}_{#1}(#2)}

\newcommandx{\alltypen}[2][2=n]{\set P_{#2}({#1})}
\newcommandx{\allcontypen}[2][2=n]{\set P_{#2}({#1})}
\newcommandx{\typeseqn}[2][2=n]{\top ^{ (#2)}({#1})}
\newcommandx{\allprob}[1]{\set P({#1})}

\let\vec\relax
\newcommand{\vec}{\mathbf}
\let\set\relax
\newcommand{\set}{\mathcal}

\newcommand{\alphabet}{\set A}
\newcommand{\alphabetX}{\set X}

\newcommand{\bfyhat}{{\mathbf{\hat  y}}}
\newcommandx{\repbfyhat}[1][1=P]{\bfyhat_{#1}}
\newcommandx{\optrepbfyhat}[1][1=P]{\bfyhat^*_{#1}}

\newcommand{\dist}{\mathsf D}
\newcommand{\cardX}{|\alphabetX|}

\newcommand{\cardA}{|\alphabet|}

\newcommand{\cardT}{|\mathcal T|}

\newcommandx{\admchannel}[1][1=\dist]{\set W^{\leq #1} }
\newcommandx{\admchanneln}[1][1=n]{\set W^{\leq \dist}_{#1} }
\newcommandx{\discontypen}[2][2=\dist]{\set W^{\leq #2}_n ({#1})}

\newcommandx{\admchannelpermn}[1][1=n]{\bar {\set W}^{\leq \dist}_{#1} }

\newcommandx{\GFp}[1][1=p]{\Field_{#1}}
\newcommandx{\Zp}[1][1=p]{\Integers_{#1}}

\newcommand{\barRo}{\bar{R}_0}
\newcommandx{\Rhv}[1][1=v]{R_{\textnormal{h},#1}}
\newcommandx{\Rv}[1][1=v]{R_{#1}}
\newcommand{\cardV}{\card{\set{V}}}
\newcommand{\cardU}{\card{\set{U}}}
\newcommand{\cardS}{\card{\set{S}}}

\begin{document}

\title{State-Dependent Channels with a Message-Cognizant Helper} 
\author{%
Amos Lapidoth, Ligong Wang, and Yiming Yan\thanks{The authors are with the Department of Information Technology and Electrical Engineering, ETH Zurich, 8092 Zurich, Switzerland (e-mail: \mbox{lapidoth@isi.ee.ethz.ch}; \mbox{ligwang@isi.ee.ethz.ch}; \mbox{yan@isi.ee.ethz.ch}). This work was supported by SNSF.}}
\date{}

\maketitle

\begin{abstract}
  The capacity of a state-dependent discrete memoryless channel
  (SD-DMC) is derived for the setting where a
  message-cognizant rate-limited helper observes the state sequence
  noncausally, produces its description, and provides the description
  to both encoder and decoder.
\end{abstract}

\section{Introduction and Main Result}

The state-dependent discrete memoryless channel (SD-DMC) is studied in
the presence of a benevolent, message-cognizant, rate-limited helper
that observes the state sequence noncausally, produces a
message-dependent rate-limited description of it, and provides the
description to both encoder and decoder. Key is that the description
may depend on the message that the encoder wishes to send. As shown in
\cite{lapidoth2023wang}, this affords the helper flexibility that may
increase capacity.

Also noteworthy is that the assistance is provided to both encoder and
decoder. Since the helper is message cognizant, it can, for example,
embed information bits in the description and in this way convey them
to the decoder error-free. There is thus some tension between using
the description bits to describe the state and using them to send data
bits directly.

Finally---unlike \cite{lapidoth2023wang}, which mostly focuses on
causal help, and \cite{rosenzweigsteinbergshamai05}, which focuses on
symbol-by-symbol help---our focus is on noncausal help. The state
sequence is thus observed by the helper noncausally, and its
description is provided to the encoder before transmission begins.

The Gaussian version of our problem was recently solved in
\cite{lapidothwangyan2023_Gauss}. For the present setting, however,
our coding scheme requires an extra element that is not present
in~\cite{lapidothwangyan2023_Gauss}.


\subsection{Problem Formulation}

Consider a SD-DMC with finite input, output, and state alphabets
$\set{X}$, $\set{Y}$, and $\set{S}$, respectively. The states
$\{S_k\}$ are IID according to $Q_S$ independently of the message $M$
to be transmitted. 
Given input $x\in\set{X}$ and state $s\in\set{S}$, the probability of
the output being $y\in\set{Y}$ is $W(y|x, s)$.

A blocklength-$n$ rate-$R$ coding scheme with rate-$\Rh$
message-cognizant assistance comprises a message set
$\M= \{1, \dots, 2^{nR}\}$; a description set
$\set{T}=\{1, \dots, 2^{n\Rh}\}$;
a helper that observes the state
sequence $S^n$ noncausally as well as the message $M$ and that
produces the description
\begin{equation}
T = h(S^n, M)
\end{equation}
using some help function
$h\colon \set{S}^{n} \times \M \to \set{T}$;
an encoder that produces the input sequence
\begin{equation}
X^n= f(M,T)
\end{equation}
using an encoding function $f\colon \M\times\set{T} \to \set{X}^n$; and a decoder that produces
the decoded message 
\begin{equation}
\hat M=\psi(Y^n, T)
\end{equation}
using some decoding function
$\psi\colon \set{Y}^{n} \times \set{T} \to \M$.

A rate $R$ is said to be achievable if there exists a sequence of
coding schemes as above (indexed by the blocklength) for which
\begin{equation}
  \lim_{n \to \infty} \Pr[\hat{M}\neq M] =  0 \label{eq:error}
\end{equation}
whenever $M$ is drawn equiprobably from $\set{M}$. 
The supremum of achievable rates is denoted $C(\Rh)$ and is the capacity we seek.

\subsection{Main Result}

Our main result is a single-letter expression of the capacity in terms
of $\Rh$.  Before formally stating the result, we introduce this
expression and some of its properties.

\begin{definition}
Define
\begin{align}
\CiRh
\triangleq \max_{\substack{Q_V, \,Q_{U|S,V},
  \,Q_{X|U,V}\colon\\I(U;S|V)\leq \Rh}} I(U;Y|V) - I(U;S|V) + \Rh \label{eq:C}
\end{align}
where $U$ and $V$ are auxiliary random variables taking values in finite sets $\set{U}$ and $\set{V}$, respectively, and where the mutual informations are computed with respect to the joint distribution
\begin{equation}
Q_V(v) \, Q_S(s) \, Q_{U|S,V}(u|s,v) \, Q_{X|U,V}(x|u,v) \, W(y|x,s).
\end{equation} 
\end{definition}

\begin{proposition}\label{prop:card}
  Without reducing the maximum in \eqref{eq:C}, $Q_{X|U,V}$ can be
  chosen to be deterministic, and the cardinalities of $\set V$ and
  $\set U$ can be restricted to $\cardV\leq 3$ and
  $\cardU\leq \cardX\cdot\cardS+1$ respectively.
\end{proposition}

\begin{IEEEproof}
See Appendix~\ref{app:card}. 
\end{IEEEproof}

\begin{proposition}
An equivalent expression for $\CiRh$ is
\begin{IEEEeqnarray}{rCl}
\CiRh &=& \max_{R_0\ge 0}  \max_{\substack{Q_V, Q_{U|S,V}, Q_{X|U,V}\colon\\I(U;S|V)\leq \Rh-R_0}} I(U;Y|V) + R_0. \label{eq:CwithR}
\end{IEEEeqnarray}
\end{proposition}
\begin{IEEEproof}
One can swap the order of the two maximizations on the RHS of \eqref{eq:CwithR} to rewrite it as
\begin{IEEEeqnarray}{c} 
\max_{\substack{Q_V, Q_{U|S,V}, Q_{X|U,V}\colon\\I(U;S|V)\leq \Rh}} \max _{0 \le R_0 \le \Rh - I(U;S|V) } I(U;Y|V) + R_0. 
\end{IEEEeqnarray}
Clearly, the inner maximum is achieved by $R_0 = \Rh - I(U;S|V)$, yielding the expression on the RHS of \eqref{eq:C}.
\end{IEEEproof}

\begin{theorem}\label{thm:C}
The capacity $C(\Rh)$ of the DMC with rate-$\Rh$ message-cognizant assistance is
\begin{equation}
C(\Rh) = \CiRh.
\end{equation}
\end{theorem}

\begin{IEEEproof}
The direct part is proved in Section~\ref{sec:achievability} and the converse part in Section~\ref{sec:converse}.
\end{IEEEproof}

\begin{remark}
  We defined the capacity under the average-probability-of-error
  criterion \eqref{eq:error}, but it is unaltered if we adopt the
  maximal-probability-of-error criterion. The only extra step in the
  code construction is to discard half the messages, specifically
  those of highest probability of error.
\end{remark}

\subsection{Discussion}

To better understand Theorem~\ref{thm:C}, we discuss below some special cases.

\bigskip

\emph{Useless channel: $W(y|x,s)$ does not depend on $x$.} For such a
channel, it always holds that $I(U;Y|V) \le I(U;S|V)$, so $\CiRh=\Rh$
(achieved by choosing $U$ to be independent of $S$). Operationally,
this capacity can be achieved by having the helper forgo describing
the state sequence and having it only send information bits to the
receiver.

\bigskip

\emph{Large assistance rate: $\Rh \ge H(S)$.} The condition $I(U;S|V)\le \Rh$ is always satisfied, so we can write $\CiRh$ as
\begin{IEEEeqnarray}{rCl}
\CiRh
&=& \max_{\substack{Q_V, Q_{U|S,V}, Q_{X|U,V}}} I(U;Y|V) - I(U;S|V) + \Rh\\
&=& \max_{\substack{Q_{U|S}, Q_{X|U}}} I(U;Y) - I(U;S) + \Rh.
\end{IEEEeqnarray}
Because the joint distribution is restricted by the Markov condition
$S\markov U\markov X$, this capacity is, in general, less than the sum
of the Gel'fand-Pinsker capacity \cite{gelfandpinsker} and $\Rh$; the
latter would correspond to a scenario where the helper can help the
encoder achieve the Gel'fand-Pinsker capacity, while at the same time
sending information bits to the receiver at the maximum possible rate
$\Rh$.

Choosing $U=(X,S)$ and $V$ deterministic yields the lower bound
\begin{IEEEeqnarray}{rCl}
\CiRh
&\geq& \max_{Q_{X|S}} I(X,S;Y) - H(S) + \Rh\\
& = & \max_{Q_{X|S}} \bigl\{ I(X;Y|S) + I(S;Y) \bigr\} + \bigl( \Rh -
H(S) \bigr) \\
& \geq & \max_{Q_{X|S}} I(X;Y|S) + \bigl( \Rh -
H(S) \bigr) 
\end{IEEEeqnarray}
with the final inequality possibly strict. Thus, even when $\Rh=H(S)$,
the capacity with a message-cognizant helper is in general larger than
with a message-oblivious helper, because the latter equals
$\max_{Q_{X|S}} I(X;Y|S)$ (namely the capacity in the absence of a
helper but when the state is known to both encoder and decoder).


\bigskip

\emph{Modulo-additive noise channel.} Consider the case where
$\set{X}=\set{Y}=\set{S} = \set A = \{0, 1, \dots, \cardA-1\}$ and,
with probability one,
\begin{IEEEeqnarray}{rCl}
Y = X\oplus S, 
\end{IEEEeqnarray}
where ``$\oplus$'' denotes the mod-$\cardA$ addition.
Using~\eqref{eq:C}, we next show that in this case
\begin{IEEEeqnarray}{rCl}
\CiRh = \log\cardA - H(S) +\Rh. 
\end{IEEEeqnarray}
We first derive an upper bound:
\begin{IEEEeqnarray}{rCl}
\CiRh &=& \max_{\substack{Q_V, Q_{U|S,V}, Q_{X|U,V}\colon\\I(U;S|V)\leq \Rh}} I(U;Y|V) - I(U;S|V) + \Rh\\
&=& \max_{\substack{Q_V, Q_{U|S,V}, Q_{X|U,V}\colon\\I(U;S|V)\leq \Rh}} I(U;X\oplus S|V) - I(U;S|V) + \Rh\\
&=& \max_{\substack{Q_V, Q_{U|S,V}, Q_{X|U,V}\colon\\I(U;S|V)\leq \Rh}} H(X\oplus S|V) - H(X\oplus S|U,V) -H(S) + H(S|U,V) + \Rh \label{eq:mod:SV} \IEEEeqnarraynumspace \\
&\leq & \log \cardA - H(S) +\Rh + \max_{\substack{Q_V, Q_{U|S,V}, Q_{X|U,V}\colon\\I(U;S|V)\leq \Rh}} H(S|U,V) - H(X\oplus S|U,V) \\
&\leq & \log \cardA - H(S) +\Rh, \label{eq:mod:sum}
\end{IEEEeqnarray}
where \eqref{eq:mod:SV} holds because $V\indep S$ (indicating that $V$
and $S$ are independent); and \eqref{eq:mod:sum} because
$X\markov (U,V) \markov S$, and because adding an independent random
variable to $S$ increases its entropy.

To obtain a lower bound, consider choosing $V$ deterministic and $U=X = B$,
where $B$ is independent of $S$ and equiprobably distributed on
$\alphabet$, i.e., $B\indep S$ and $B \sim\Unif(\alphabet)$. This yields 
\begin{IEEEeqnarray}{rCl}
\CiRh&\geq&  I(U;Y) - I(U;S) + \Rh\\
&= & I(B; B \oplus S) - I(B;S) + \Rh\\
&=& H(B \oplus S) - H(B \oplus S|B) + \Rh\\
&=&  \log \cardA  - H(S)+ \Rh,
\end{IEEEeqnarray}
where, in the last step, we used the fact that $B \oplus S$ is uniform
over $\set{A}$ (because $B$ is uniform), and that
$ H(B \oplus S|B) = H(S|B)=H(S)$.

In fact, this capacity can also be achieved with the simple scheme in
which the helper ignores the states and uses its entire rate $\Rh$ to
send information bits to the receiver.


%


%
%

\section{Achievability}\label{sec:achievability}

We first show how to achieve any rate below
\begin{equation}
\max_{\substack{Q_{U|S},Q_{X|U}\colon \\ I(U;S)\le \Rh}} I(U;Y). \label{eq:IUY}
\end{equation}
This and a time-sharing argument will then imply the achievability of any rate below $\CiRh$.
%
%
%
%
By the convexity of $I(U;Y)$ in the conditional distribution of $Y$
given $U$, the maximum in \eqref{eq:IUY} is unaltered when we restrict
$Q_{X|U}$ to be deterministic. It thus suffices to prove the
achievability of $I(U;Y)$ when $Q_{U|S}$ is such that $I(U;S)<\Rh$ and
for $X = \phi(U)$ for some arbitrary mapping
$\phi \colon \set U \to \set
X$. 

Generate $2^{n(R+\Rh)}$ length-$n$ codewords
$\{u^n(m, t)\}_{(m,t)\in\M\times\T}$ independently, with the $n$
components of each codeword being drawn IID according to the marginal
distribution $Q_U$.  If the message to be transmitted is $m\in\M$, and
if it observes the channel state sequence $s^{n}$, the helper searches
the $2^{n\Rh}$ codewords $\{u^n(m,t)\}_{t \in \set{T}}$ for some
$u^n(m,t^{\star})$ that is strongly jointly typical with~$s^{n}$ with
respect to the $(U,S)$-marginal $Q_{U,S}$. The helper is very likely
to find such a~$t^{\star}$ because $I(U;S)<\Rh$. Having found
$t^{\star}$, the helper reveals it to the encoder and the decoder. The
encoder now transmits $\phi(u^n(m,t^{\star}))$, where $\phi(\cdot)$
extends to $n$-tuples component-wise.  The decoder, with $t^{\star}$
and the received sequence $y^{n}$ in hand, searches for a some
$\hat{m}$ for which $u^n(\hat{m},t^{\star})$ is strongly jointly
typical with 
$y^{n}$ with respect to the $(U,Y)$-marginal $Q_{U,Y}$.

We next analyze the scheme. Whenever the helper succeeds in finding a
suitable $t^\star$ (which, we recall, happens with high probability),
the transmitted codeword $u^n$ is strongly jointly typical with
$s^n$. In this case, since $x^n$ is a deterministic function of $u^n$,
the triple $(u^n, s^n, x^n)$ must also be strongly jointly
typical. Since $U\markov (X,S) \markov Y$ forms a Markov chain under
the joint PMF at hand, and since $Y^n$ is the channel response to
$(x^n, s^n)$, it follows from the Markov Lemma \cite[Lemma
12.1]{gamalkimtextbook}, \cite[Corollary 4.12]{moser2023advanced} that,
with high probability, $Y^n$ is strongly jointly typical
with~$u^n$.

As to an incorrect codeword, say $u^n(m',t^\star)$ for some
$m'\neq m$, it
is drawn independently of~$Y^{n}$, so the probability that it is
jointly typical with $Y^n$ is approximately $2^{-n I(U;Y)}$. Since
there are $(2^{nR}-1)$ such codewords, 
the average error probability will vanish with the blocklength
whenever $R<I(U;Y)$. This establishes the achievability of all rates
smaller than \eqref{eq:IUY}.

Next, we introduce a time-sharing random variable $V$. For each of its
realizations, say $V=v$, we construct a coding scheme as above with
transmission rate $R_v$ and assistance rate $\Rhv$, to be used
in $n \cdot Q_V(v)$ of the $n$ channel uses in a block. This
allows us to achieve all rates below
\begin{equation}
\max_{\substack{Q_V,Q_{U|S},Q_{X|U,S}\colon \\ I(U;S|V)\le \Rh}} I(U;Y|V).
\end{equation}

Finally, we introduce $R_0$, which corresponds to the rate at which
the helper sends information bits directly to the
receiver. Specifically, the assistance $T$ is a tuple
$(T_1,\tilde{M})$, where $T_1$ is of rate $\Rh-R_0$, which is substituted
for the help rate in the scheme above (with the joint distribution now
satisfying $I(U;S|V)\leq \Rh -R_0$), while $\tilde{M}$ is of rate
$R_0$, and is part of the message to be communicated. The total
transmission rate is then $I(U;Y|V) + R_0$. Optimizing over $R_0$ and
the distributions yields the achievability of the RHS of
\eqref{eq:CwithR} and establishes the direct part of
Theorem~\ref{thm:C}.



\section{Converse}\label{sec:converse}

Consider a message $M$ that is drawn equiprobably from $\M$. Fano's
inequality implies that, for any sequence of rate-$R$ coding schemes
with rate-$\Rh$ message-cognizant assistance and vanishing average
probability of error, there exists some sequence $\{\delta_n\}$
tending to zero such that
\begin{IEEEeqnarray}{rCl}
nR-n\delta_n 
 &\le & I(M; Y^n ,T)\\ 
 &=& I(M;Y^n|T)+I(M;T)\\
 &=& H(Y^n|T)-H(Y^n|M,T)+I(M;T)\\
 &=& \sum_{k=1}^n H(Y_k|T,Y^{k-1})-H(Y_k|M, T,Y^{k-1}) + I(M;T)\\
 &\leq& \sum_{k=1}^n H(Y_k)-H(Y_k|M, T,Y^{k-1}) + I(M;T)\\
 &=&  \sum_{k=1}^n I(Y_k; U_k) + I(M;T), \label{eq:Uk}\\
 &=&  n \, I(Y_V; U_V | V) + n \, \barRo,\label{eq:YVUV}
\end{IEEEeqnarray}
where in \eqref{eq:Uk} we introduce $U_k \defn(M, T,Y^{k-1})$ for
$k\in[n]$; and in \eqref{eq:YVUV} we define
$\barRo\defn \frac{1}{n}I(M; T)$, and introduce an auxiliary random
variable $V$ that is equiprobable over $[n]$ and is independent of all
other random quantities. Dividing by $n$,
\begin{IEEEeqnarray}{rCl}
R- \delta_n &\leq &  I(Y_V; U_V | V) + \barRo .\label{eq:upperI}
\end{IEEEeqnarray}
To relate this upper bound to \eqref{eq:CwithR}, we proceed to justify that, by our choice of  $V$ and $U_V$, the joint distribution of $(V, U_V, X_V, S_V, Y_V)$ satisfies the conditions in the  maximization in  \eqref{eq:CwithR}. 

First, we verify that the joint distribution factorizes correctly as 
\begin{IEEEeqnarray}{rCl}
Q_V \circ Q_{S_V}\circ Q_{U_V|S_V,V} \circ Q_{X_V|U_V,V}\circ Q_{Y_V|X_V ,S_V}.
\end{IEEEeqnarray}
Because $\{S_k\}$ are IID $\sim Q_S$, we have $S_V \indep  V$. The Markov relation $S_V\markov ( U_V,V) \markov X_V$ holds because $X_V$, as a function of $(M, T, V)$, is a function of $( U_V,V)$. Finally, the Markov relation  $ (V,U_V) \markov (X_V,S_V) \markov Y_V$ holds 
because the channel is memoryless. 

We next show that $ I(S_V; U_V | V) \le \Rh-\barRo$ as follows:. 
\begin{IEEEeqnarray}{rCl}
n \, I(S_V; U_V | V) &=& \sum_{k=1}^n I(S_k; U_k)\\
 &=& \sum_{k=1}^n I(S_k; M, T,Y^{k-1})\\
 &\leq& \sum_{k=1}^n I(S_k; M, T,Y^{k-1}, S^{k-1})\\
 &=& \sum_{k=1}^n I(S_k; M, T, S^{k-1})\label{eq:eliminateXY}\\
 &=& \sum_{k=1}^n I(S_k; M, T|S^{k-1})\\
 &=& I(S^n; M, T)\\
 &=& I(S^n; T|M)\\
 & = & I(S^n,M; T) - I(M; T) \\
  &\leq&  \log\cardT - n \barRo\\
 &=& n \left(\Rh - \barRo\right), 
\end{IEEEeqnarray}
where \eqref{eq:eliminateXY} holds because, given $(M, T, S^{k-1})$,
the past channel inputs $X^{k-1}$ are deterministic, so $Y^{k-1}$ is a
function only of the channel randomness and is thus independent of all other
random variables.
%

Finally, letting $n$ tend to infinity yields the desired converse.

\appendix

\section{Proof of Proposition~\ref{prop:card}}\label{app:card}

To see why the maximum in~\eqref{eq:C} is unaltered when $Q_{X|U,V}$
is restricted to be deterministic, we fix
$Q_S, Q_V, Q_{U|S,V}, Q_{Y|X,S}$ and note that this fixes
$ I(U;S|V)$ and causes $I(U;Y|V) $ to be a function of $Q_{X|U,V}$
only. Since $I(U;Y|V) $ is convex in $Q_{Y|U,V}$, and since the latter
is linear in $Q_{X|U,V}$, it follows that $I(U;Y|V) $ is convex in $Q_{X|U,V}$,
and thus establishes that the maximization can be achieved with $Q_{X|U,V}$ being
deterministic.

To derive cardinality bounds on $\set V$ and $\set U$, we first rewrite \eqref{eq:C} as
\begin{align}
\CiRh = \max_{\substack{Q_V,\, \{\Rhv\}_{v\in\set V}\colon \\ \E{\Rhv[V] }= \Rh}} 
\sum_{v\in\set V} Q_V(v) \left(\max_{\substack{Q_{U|S}, Q_{X|U}\colon\\I(U;S)\leq \Rhv}} I(U;Y) - I(U;S) + \Rh\right).\label{eq:Csplit}
\end{align}

To find an upper bound on $\set U$, we fix  $\Rhv$ and focus on the parenthesized term in \eqref{eq:Csplit}. We follow the same technique as in \cite[Appendix C]{gamalkimtextbook}. 
Fix any $U\in \set{U}$ such that  $S\markov U\markov X$. Consider the set $\P_{\times}$ of all product PMFs on $\set X\times \set S$, which is connected and compact, and, by the Markov relation, contains $Q_{X,S|U=u}$ for all $u\in\set U$. The following $(\cardX\cdot \cardS+1)$ functions of $\pi\in \P_{\times}$ are all continuous: 
\begin{IEEEeqnarray}{rCl}
g_1(\pi) &=& H(Y),\\
g_2(\pi) &=& H(S),\\
g_{i,j}(\pi) &=& \pi(i, j), \quad (i,j)\in \set X\times \set S\setminus\{(\cardX, \cardS)\}, 
\end{IEEEeqnarray}
where the continuity of $g_1(\cdot)$ and $g_2(\cdot)$ follows from the continuity of entropy and the linearity of $Q_Y$ in $\pi$. Applying the support lemma \cite[Appendix C]{gamalkimtextbook}, we can find a random variable $U'$, 
taking at most  $(\cardX\cdot \cardS+1)$ values in $\set U'\subseteq \set U$, such that, when substituting $U'$ for $U$, the values of 
$H(Y|U')$ and $H(S|U')$ and the distribution $Q_{X,S}$ are preserved. The latter implies that $Q_{S}$ and $Q_Y$ are preserved, and hence also $H(Y)$ and $H(S)$. To sum up, it is therefore sufficient to restrict $\cardU\leq \cardX\cdot\cardS+1$, while preserving the values of $Q_{S}$, $I(U;Y)$, and $I(U;S)$, and preserving the Markov relation $S\markov U\markov X$.

To prove the desired upper bound on $\set V$, let $g(\Rhv)$ denote the term in the parenthesis of \eqref{eq:Csplit}. Given any choice of $\set V$, $Q_V$, and $\{\Rhv\}_{v\in\set V}$, we write the two-dimensional vector $\big(\Rh,\, \sum_{v\in\set V} Q_V(v)g(\Rhv)\big)$ as the convex combination
\begin{IEEEeqnarray}{rCl}
\begin{pmatrix}
\Rh\\ \sum_{v\in\set V} Q_V(v)g(\Rhv)
\end{pmatrix}
= \sum_{v\in\set V} Q_V(v)
\begin{pmatrix}
\Rhv\\g(\Rhv)
\end{pmatrix}.
\end{IEEEeqnarray}
Applying Carath\'eodory's Theorem \cite[Appendix A]{gamalkimtextbook}, we obtain that it is sufficient to have $\cardV\leq 3$. 

\bibliographystyle{hieeetr}
\bibliography{./bibliography_gauss.bib}

\end{document}